\documentclass[square]{ws-procs11x85}
\usepackage{balance} 

\def\Journal#1#2#3#4{{#1} {\bf #2}, #3 (#4)}

\newcommand{\met}{\hbox{E\kern-0.5em\lower-0.1ex\hbox{/}}_T}

\def\ginj{\Gamma_{\rm e}}
\def\etab{\eta_{\mathrm{B}}}
\def\tbb{T_{\mathrm{BB}}}

\begin{document}

\twocolumn[
\title{Electron thermalization and photon emission from magnetized compact sources}

\author{Indrek Vurm$^{1,2}$ and Juri Poutanen$^{1}$}

\address{$^{1}$ Astronomy Division, Department of Physical Sciences, P.O.Box 3000, 90014 University of Oulu, Finland \\E-mail: indrek.vurm@oulu.fi,
 juri.poutanen@oulu.fi}

\address{$^{2}$ Tartu Observatory, 61602 T\~{o}ravere, Tartumaa, Estonia} 


\begin{abstract}
We present detailed calculations of the electron thermalization by synchrotron self-absorption accounting for cooling by Compton scattering. For the first time, we solve coupled kinetic equations for electrons and photons without any approximations on the relevant cross-sections and compute self-consistently the resulting electron and photon distributions.
The results presented in this contribution may be applied to the magnetized coronae around Galactic black hole accretion disks.
\end{abstract}
\vskip12pt  
]

\bodymatter

\section{Introduction} 

We have developed a numerical code  to study radiative processes in relativistic magnetized plasmas.
Previous works made various simplifications making them useful only in certain parameter regimes.
Usually one assumes that Compton scattering takes place in the Thomson regime.
Often one treats all electrons as ultrarelativistic and neglects diffusive thermalizing processes
such as synchrotron self-absorption operating at lower energies. In some works the kinetic equations are not solved
self-consistently (see e.g.  \cite{GHS}), instead some assumptions are invoked for the distribution of
one or the other type of particles.
One of the well-known codes dealing with radiative processes in similar conditions is {\sc eqpair} by Coppi \cite{Coppi},
which however does not consider synchrotron thermalization, but instead treats Coulomb collisions as the
thermalizing process. 

The only major simplifying assumption in our treatment concerns the geometry of the active region.
Our calculations are done in a simple one-zone geometry with a tangled
magnetic field and isotropic particle distributions.
However, we have tried to avoid the other approximations mentioned above by accurately treating the microprocesses involved.
We have used the exact Klein-Nishina Compton scattering cross-sections \cite{NP94} 
for both types of particles and exact cyclo-synchrotron emissivities. 
The coupled time-dependent kinetic equations for electrons and photons are solved self-consistently,
making no a priori assumptions about the distributions of either type of particles.
Such approach makes our code applicable in a wide range of parameters.

\begin{figure*}[t]
\begin{center}
\leavevmode \epsfxsize=6.0cm \epsfbox{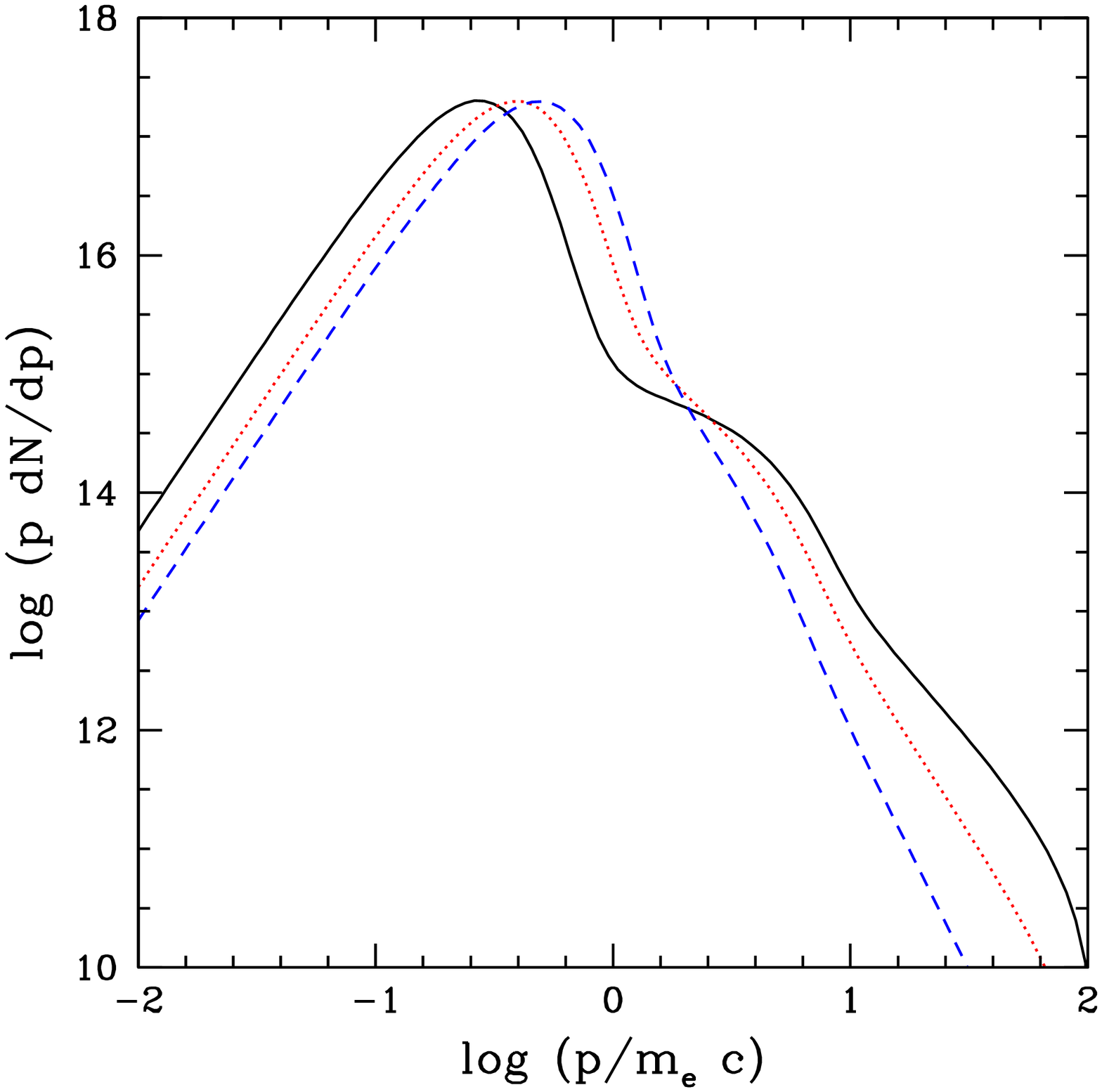} \hspace{1cm}
\epsfxsize=6.0cm \epsfbox{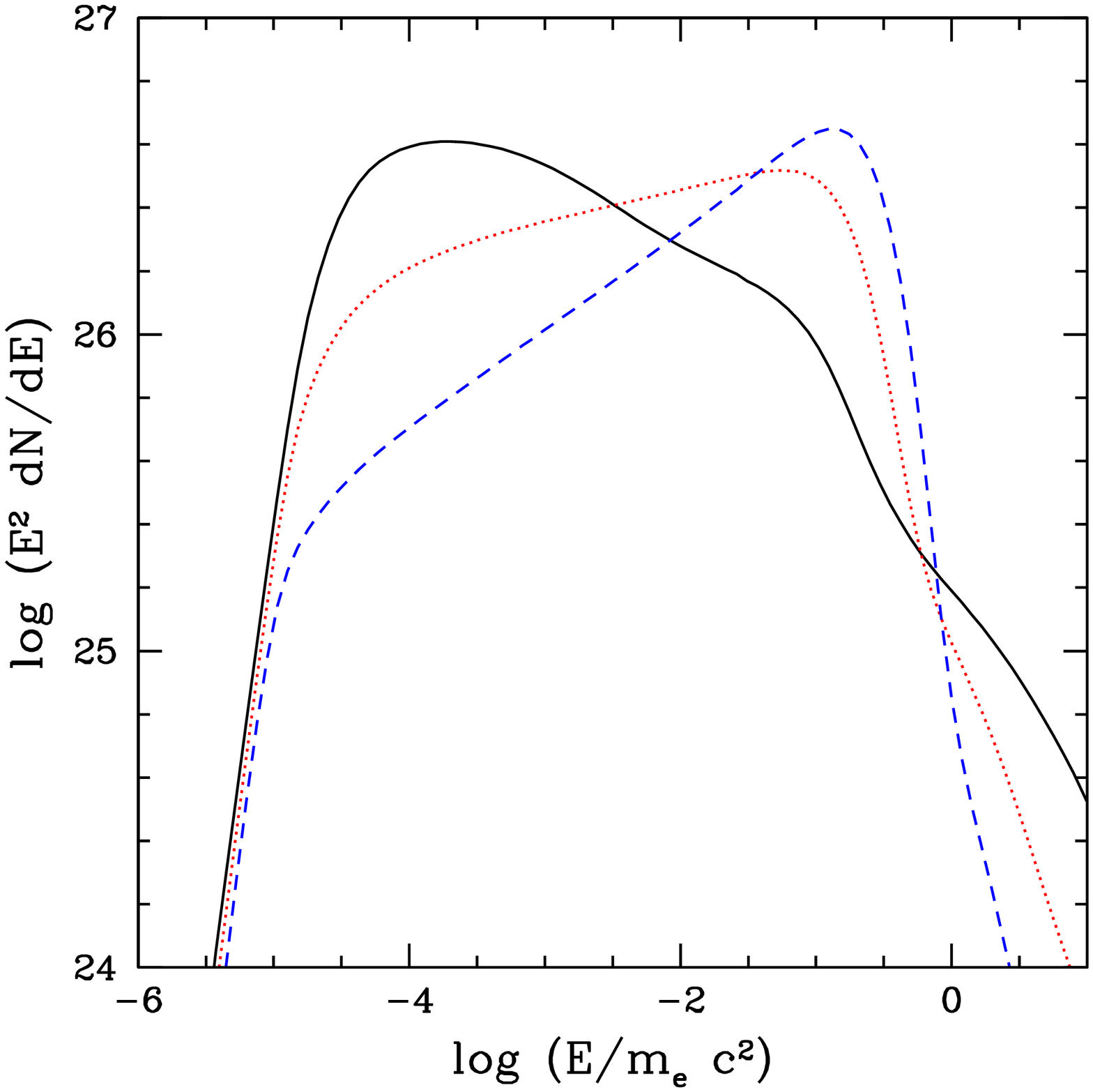}
\end{center}
\caption{{Equilibrium electron (left) and photon (right) distributions 
for different injection slopes: $\ginj=2$ (solid line), $3$ (dotted line) and $4$ (dashed line).
Other parameters: $\tau_\mathrm{T}=2$, $L=10^{37}$ erg s$^{-1}$, $\etab = 5$, no external radiation ($f = 0$). The Maxwellian parts of the distributions correspond
to electron temperatures of $12$, $24$ and $36$ keV, respectively.
}		
}	
\label{fig1}
\end{figure*}

\section{Code description} 

We solve numerically coupled kinetic equations for photons  and electrons: 
\begin{align}
\frac{\partial f_{\mathrm{ph}}}{\partial t} = \frac{Df_{\mathrm{ph}}}{Dt}_{\mathrm{S}} +\frac{Df_{\mathrm{ph}}}{Dt}_{\mathrm{C}}
- \frac{f_{\mathrm{ph}}}{t_{\mathrm{ph, esc}}} + Q_{\mathrm{ph}},			\label{eq1}
\end{align}
\begin{align}
\frac{\partial f_{\mathrm{e}}}{\partial t} = \frac{Df_{\mathrm{e}}}{Dt}_{\mathrm{S}} +\frac{Df_{\mathrm{e}}}{Dt}_{\mathrm{C}}
- \frac{f_{\mathrm{e}}}{t_{\mathrm{esc}}} + Q_{\mathrm{e}}.			\label{eq2}
\end{align}
Here the quantities $Df/Dt$ describe the rates of different processes (C -- Compton scattering, S -- synchrotron) 
and contain both integral and differential terms.
The quantities $Q_{\mathrm{ph}}$ and $Q_{\mathrm{e}}$ denote sources of particles not associated with synchrotron or Compton processes, such as
injection of high-energy electrons and external sources (e.g. disk) of photons.

The treatment of Compton scattering over a wide energy range is not straightforward due to the
very different behaviour of the process in different regimes. We have therefore adopted an approach, where we treat the process
as discrete or alternatively as a continuous energy gain/loss mechanism depending on whether or not the redistribution function
can be resolved by our grid. This results in the presence of both integral and up to second order differential terms
(to account for diffusion/thermalization) in $Df/Dt_{\mathrm{C}}$ for both photons and electrons.
To account for synchrotron cooling and thermalization, the rate $Df_{\mathrm{e}}/Dt_{\mathrm{S}}$ also contains up to second order derivative terms.
The resulting integro-differential equations (\ref{eq1}) and (\ref{eq2}) are discretized on energy and time grids and solved
iteratively as two coupled systems of linear algebraic equations.

\section{Setup}

To demonstrate some of the capabilities of our code we have arranged a simple setup, where
we inject a power-law distribution of electrons into the emission region permeated by the magnetic field and let them
cool and thermalize by synchrotron and Compton processes. The seed photons for Compton upscattering are produced by
synchrotron emission, in the last figures we also add an external source of soft disk photons.
The main parameters are as follows: \\
a) $\ginj$: injection slope of the power-law, $\gamma^{-\ginj}$, electrons,
extending from $\gamma=1$ to $100$ in all cases.\\
b) $\tau_\mathrm{T}$: equilibrium Thomson optical thickness. \\
c) $R$: size of the emission region, fixed at $3\times 10^7$ cm. \\
d) $L$: total power entering the active region as electron injection and external soft radiation. \\
e) $f$: ratio of external soft photon compactness (dimensionless ratio $L/R$)
and injected electron compactness, $l_{\mathrm{disk}}/l_{\mathrm{inj}}$. \\
f) $\etab$: ratio between magnetic and electron injection compactnesses. \\
g) $\tbb$: disk blackbody temperature (not a free parameter, but depends on $L$ and $R$). \\

\section{Numerical results}

We now present the results of simulations for different parameters 
as equilibrium electron and photon distributions.

\begin{figure*}[t]
\begin{center}
\leavevmode \epsfxsize=6.0cm \epsfbox{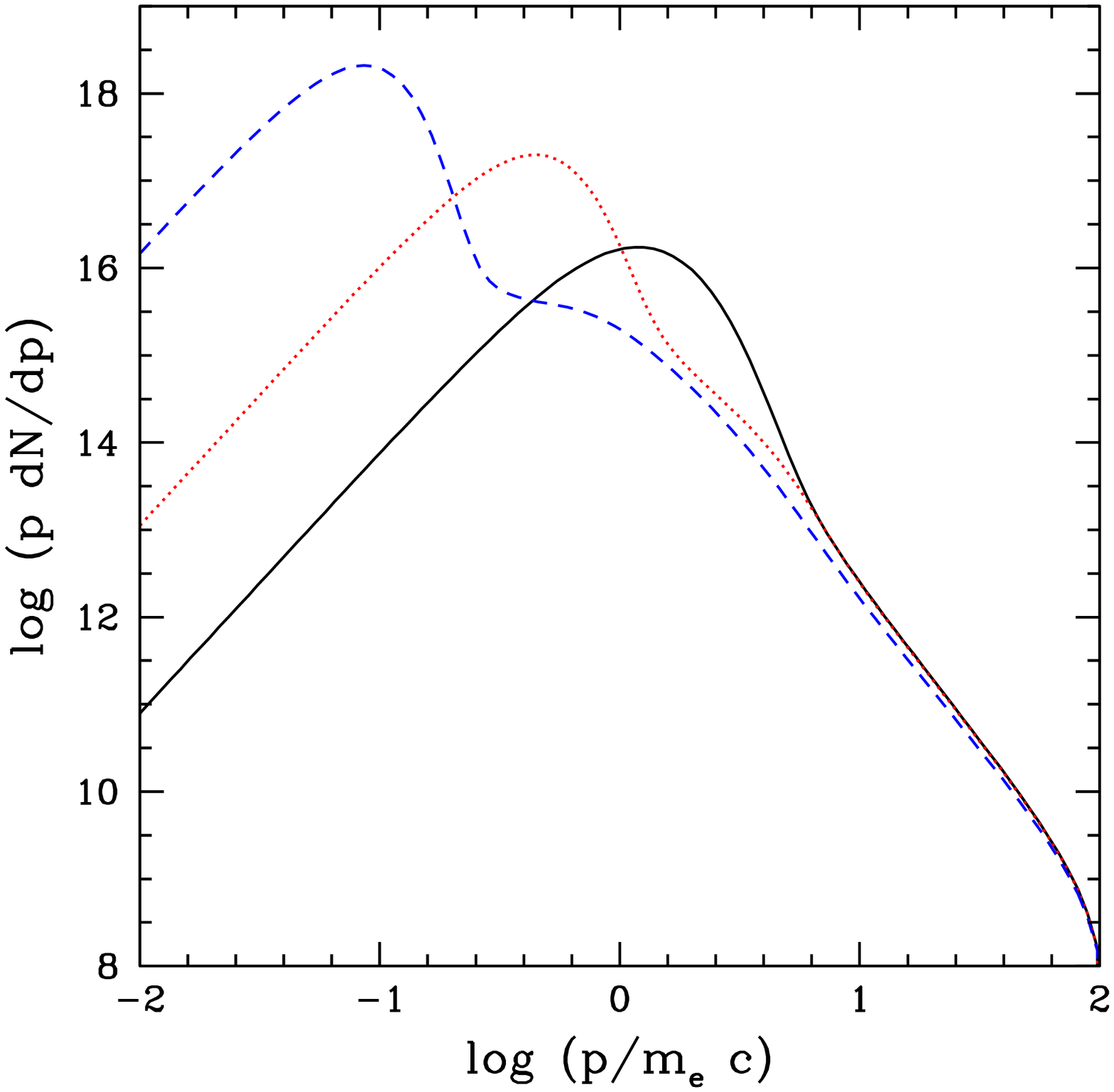} \hspace{1cm}
\epsfxsize=6.0cm \epsfbox{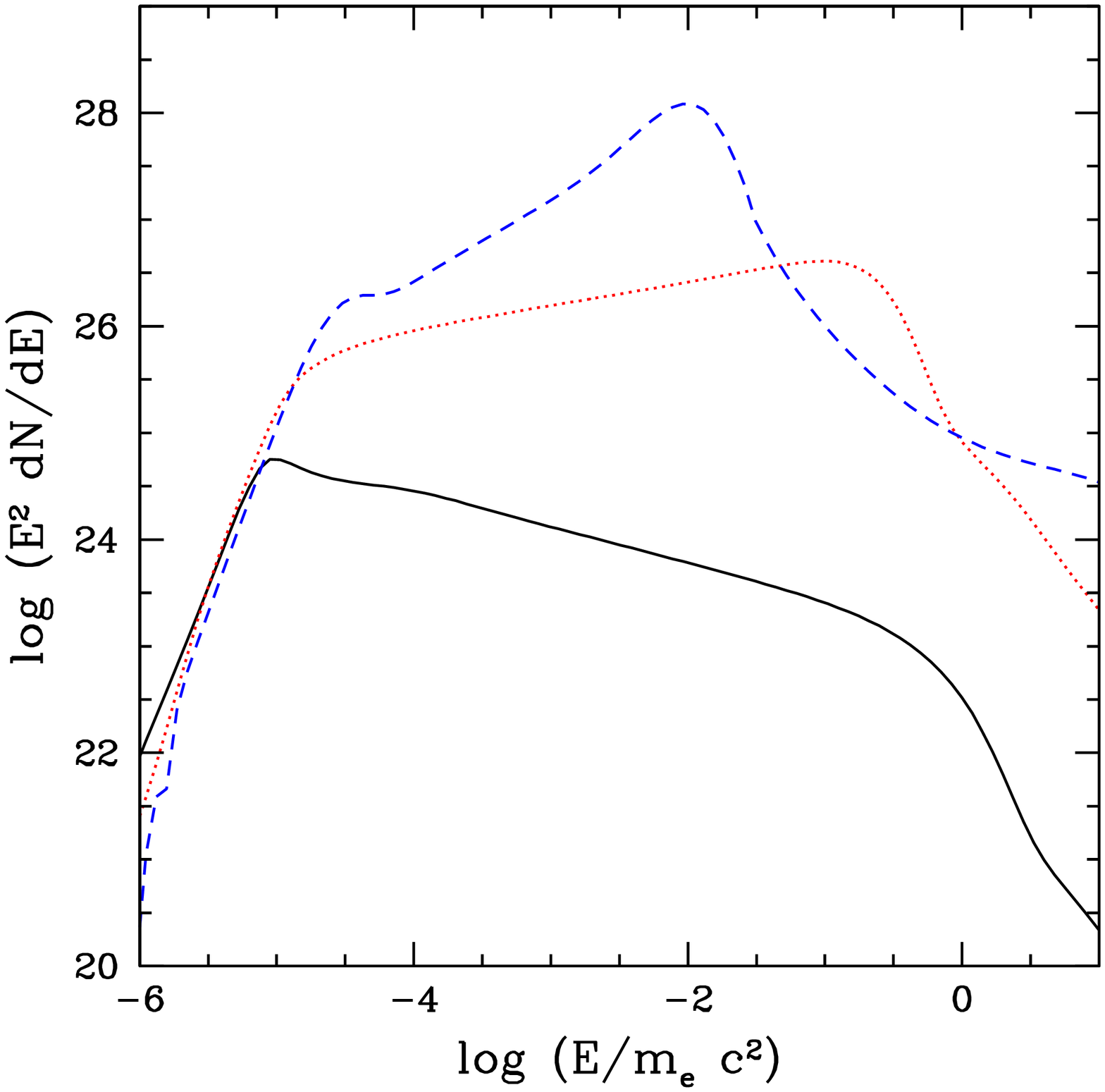}
\end{center}
\caption{{Equilibrium electron (left) and photon (right) distributions 
for varying injection power: $L=10^{36}$ (solid line), $10^{37}$ (dotted line), $10^{38}$ erg s$^{-1}$ (dashed line).
The electron escape timescale is held fixed, resulting in different optical thicknesses in the steady-state:
$\tau_\mathrm{T}=0.2$ (solid), 2 (dotted), 20 (dashed), respectively, with corresponding 
equilibrium electron temperatures of 140, 30 and 1.3 keV.
Other parameters: $\ginj=3.5$, $\etab=5$, no external radiation.
}		
}	
\label{fig2}
\end{figure*}

\begin{figure*}[t]
\begin{center}
\leavevmode \epsfxsize=6.0cm \epsfbox{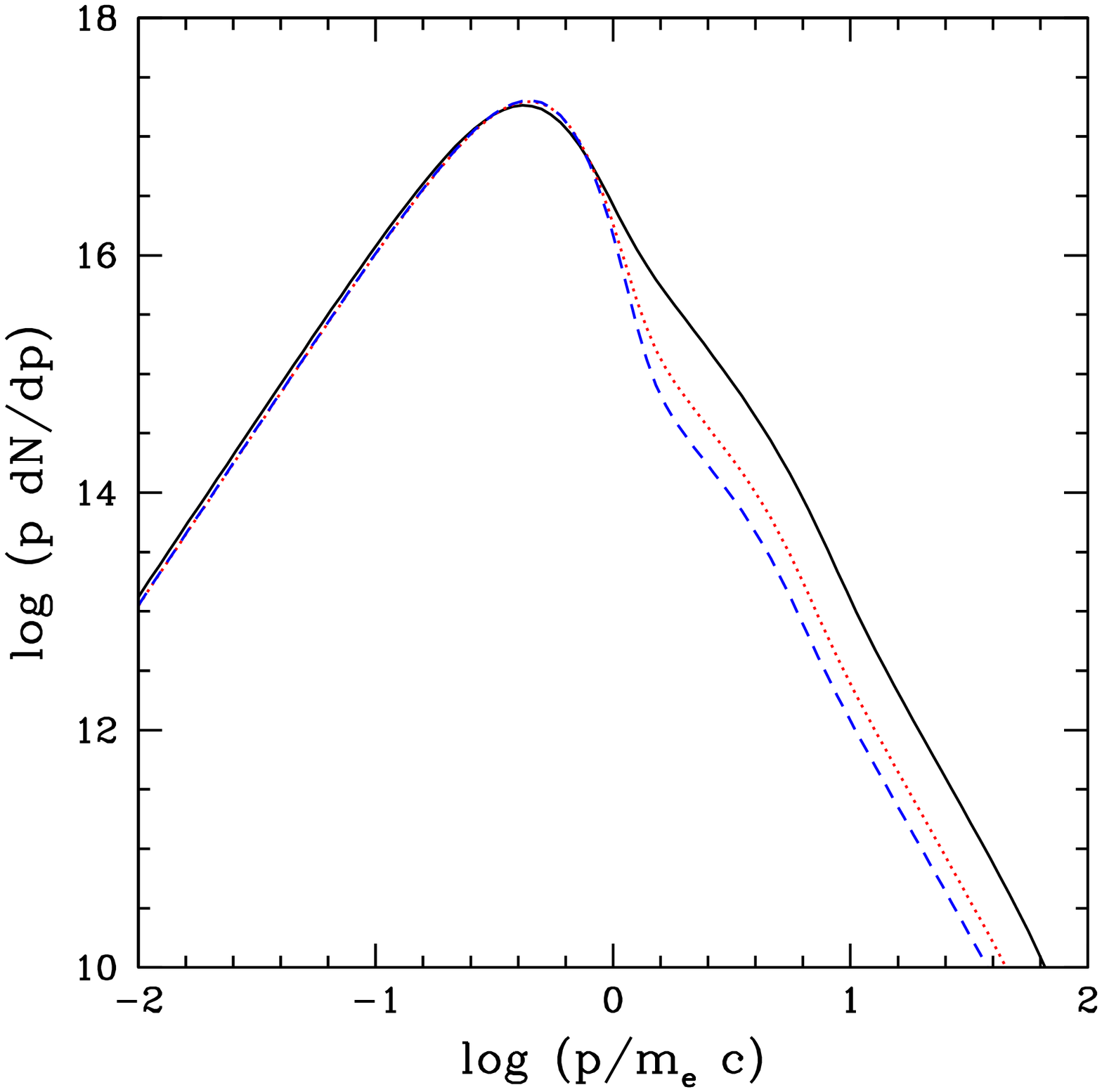} \hspace{1cm}
\epsfxsize=6.0cm \epsfbox{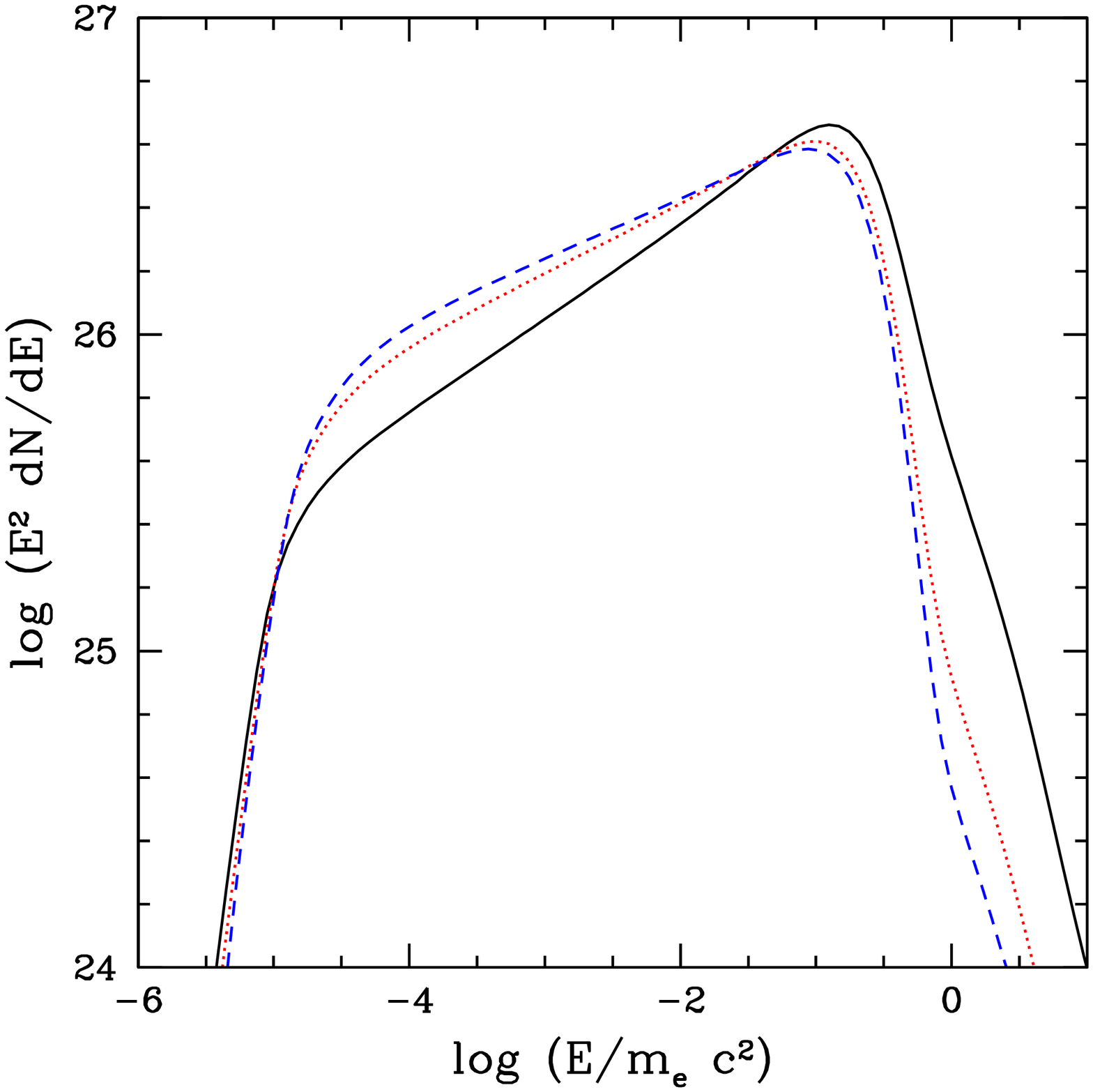}
\end{center}
\caption{{Equilibrium electron (left) and photon (right) distributions
 for variable $\etab$: $\etab = 1$ (solid), 5 (dotted),  10 (dashed).
Other parameters: $\ginj=3.5$, $\tau_\mathrm{T}=2$, $L=10^{37}$ erg s$^{-1}$, no external radiation.
Equilibrium electron temperatures are around $30$ keV.}		
}	
\label{fig3}
\end{figure*}

\subsection{Variable injection slope}

The low-energy electrons absorb synchrotron radiation produced by the high-energy tail of the electron distribution. 
The more power is in the nonthermal tail (the flatter the tail) the higher is the self-absorption frequency and the 
more seed photons there are for Compton upscattering, in accordance with \cite{WZ01}.
Note that a larger amount of seed photons
results in a drop of the electron temperature (Fig. \ref{fig1}, left). Also take note of the curious secondary "bump" that develops 
in case of relatively flat injection spectrum. It is caused by the fact that synchrotron emission produced by these
electrons is still strongly self-absorbed, while the energy losses and gains stay close to each other for an extended energy interval.
In fact, from simple arguments one can show that the ratio of energy loss and gain rates for relativistic power-law electrons emitting
in self-absorbed regime is approximately
\begin{align}
\frac{\dot{\gamma}_\mathrm{c}}{\dot{\gamma}_\mathrm{h}} = \frac{5}{p + 2},		\label{eq3}
\end{align}
where $p$ is the index of the steady-state power-law electron distribution.
Observe that for $p=3$ the heating and cooling rates
are balanced. However, as discussed by Rees\cite{Rees} already in 1967 and as can be seen from (\ref{eq3}),
such quasi-equilibrium is unstable.  

As we see from Fig. \ref{fig1}, the Comptonized spectrum for hard injection $\ginj=2$ is much softer than the spectra observed in
the hard state of Galactic black holes (GBH), even without any contribution to the cooling by the disk. This strongly constrains
the electron injection mechanism in these sources.

\subsection{Variable injected power}

Emission at the synchrotron self-absorption frequency is dominated by the
thermal part of the electron distribution in case of low injection rate/optical thickness (solid line in Fig. {\ref{fig2}}),
but it is dominated by the power-law tail in the cases of higher luminosity. 
As the injection rate increases, so does the number of seed photons for upscattering and the equilibrium 
electron temperature drops. The Compton $y$-parameter also increases 
despite the decreasing electron temperature and the spectrum hardens, showing signs of becoming saturated at high $L$.

\begin{figure*}[t]
\begin{center}
\leavevmode \epsfxsize=6.0cm \epsfbox{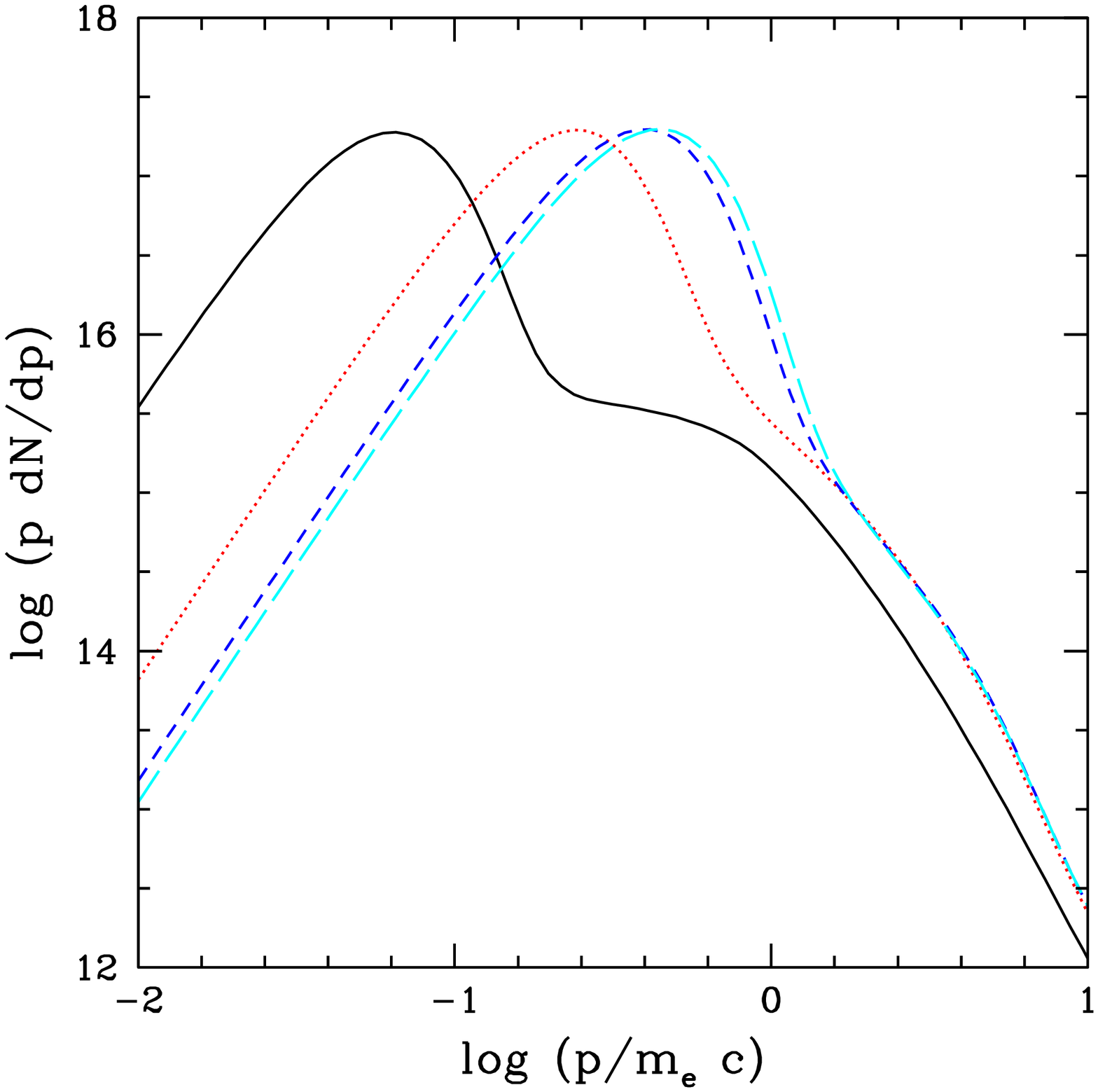} \hspace{1cm}
\epsfxsize=6.0cm \epsfbox{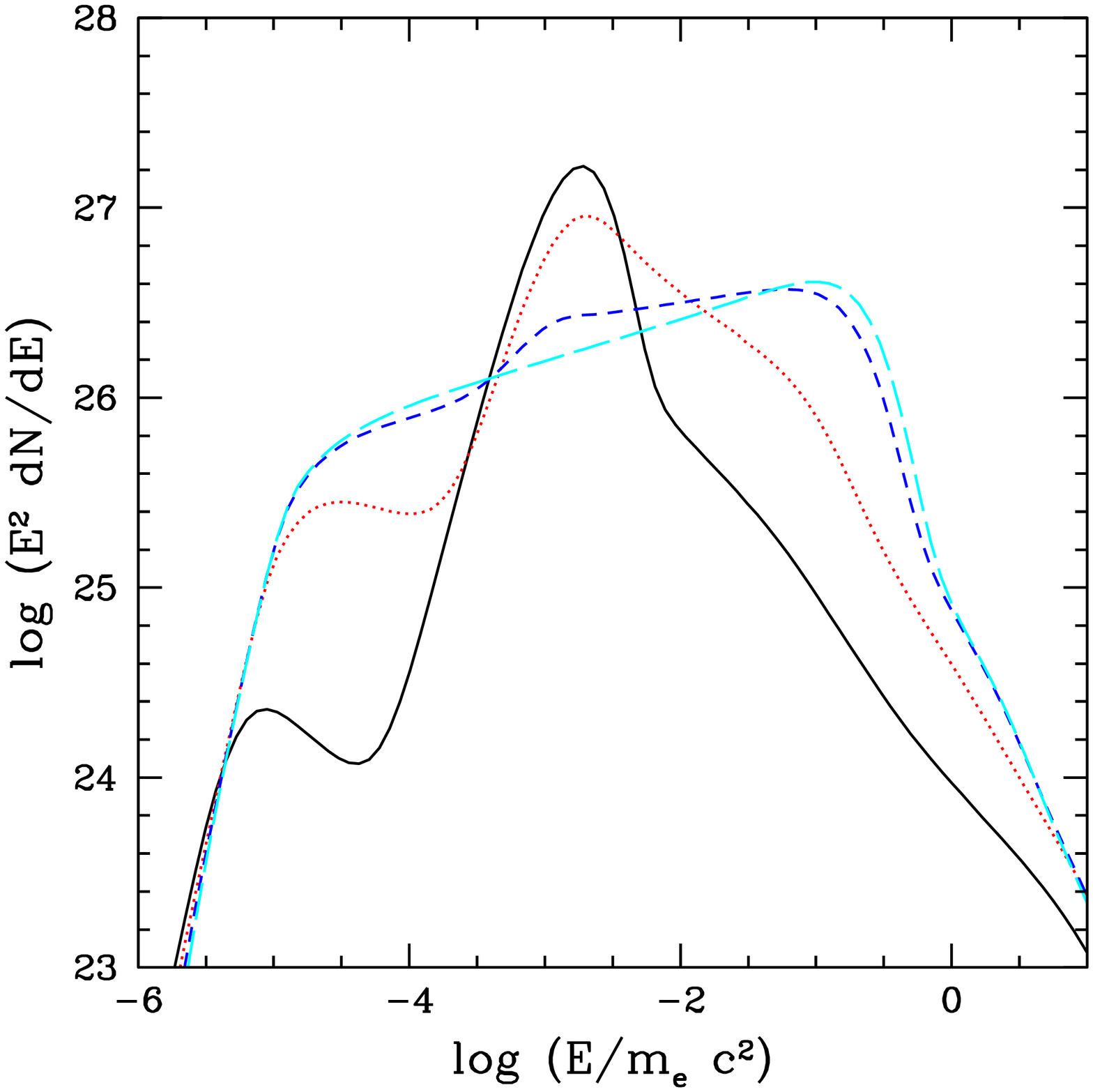}
\end{center}
\caption{Equilibrium electron (left) and photon (right) distributions for different ratios between the power from external
soft photons (disk) and the injected power: $f = 0$ (long dashed), 0.1 (short dashed), 1 (dotted), 10 (solid).
Other parameters: $\ginj=3.5$, $\tau_{\mathrm{T}}=2$, $L=10^{37}$ erg s$^{-1}$, $\etab=5$. 
The transition from $f=0$ to $f=10$ affects the observed spectrum
and strongly resembles the spectral state transitions in GBH.}	
\label{fig4}
\end{figure*}

\subsection{Variable ratio between magnetic and injected compactnesses}

Increasing magnetic compactness increases the electron cooling rate, therefore the normalization of the cooling power-law decreases, while the thermal distribution persists to higher energies (Fig. {\ref{fig3}}). The synchrotron radiation near the self-absorption frequency is produced by the power-law tail of the electron distribution in all three cases. The photon energy index $\alpha$  of the thermal Comptonization spectrum $\frac{{\rm d}N}{{\rm d}E}$$\propto$$ E^{-(\alpha+1)}$ varies between 0.7 
and~0.8.

\subsection{Variable ratio between disk and injected power}

Increasing the ratio between the soft disk luminosity and the injected power
leads to the evolution from almost purely thermal to a hybrid (mostly non-thermal) electron distribution (Fig. {\ref{fig4}}).
The resulting photon distribution changes from hard thermal Compton dominated
spectrum to the one dominated by the disk blackbody, with a weak non-thermal tail. This is similar to what is observed in the
spectral state transitions in GBH \cite{PC98}.

\section{Conclusions}

1. We have shown that hard electron injection is unable to produce hard Comptonized spectra
because the resulting high synchrotron luminosity provides too many soft seed photons for Compton cooling of the electrons. We are therefore able to
constrain the electron injection mechanism operating in the hard state of GBH. \\
2. A source at $L=10^{37}$ erg s$^{-1}$, where only synchrotron and Compton processes operate, produces an X-ray spectrum with
$\alpha\approx 0.7$, which strongly resembles the hard state of GBH. The feedback from the disk does not seem to be needed to produce such spectra (in contrast to \cite{HM93,S95,MBP01}). \\
3. At high luminosities $L\sim 10^{38}$ erg s$^{-1}$, in the absence of disk radiation, the spectrum is close to saturated Comptonization,
peaking at $\sim 5$ keV. This Wien-type spectrum might be associated with the thermally-looking emission in the
very high states of e.g. GRS 1915+105.\\
4. A behaviour similar to what is observed in the spectral state transitions in GBH can be reproduced
by varying the ratio of injected soft luminosity and the power dissipated in the corona.

\section*{Acknowledgments}
 
The work was supported by the CIMO grant TM-06-4630 and 
the Academy of Finland grants 122055 and 112982.

\end{document}